\documentclass[12pt,twoside]{article}

\def\TheAuthor{Stephan D\"{u}rr}                             
\def\TheTitle{Introduction to Physical Implementations of Quantum Information Processing\footnotemark}
\def\TheHeading{Physical Implementations of QIP}             
\def\TheInstitute{Max-Planck-Institut f\"{u}r Quantenoptik}  
\def\TheAddress{}                                            
\def\ThePart{A}                                              
\def\TheChapter{3}                                           

\usepackage{ferienschule_12}                                  

\usepackage{url}
\newcommand{\lra}[2]{|{#1}\rangle\leftrightarrow|{#2}\rangle}
\newcommand{\ke}[1]{|#1\rangle}
\newcommand{\kb}[2]{|#1\rangle\langle#2|}
\newcommand{\lr}[1]{\langle#1\rangle}
\renewcommand{\Re}{\operatorname{Re}}
\renewcommand{\Im}{\operatorname{Im}}

\begin{document}

\MakeTitel           

\tableofcontents     

\footnotetext{Lecture Notes of the $44^{{\rm th}}$ IFF Spring
School ``Quantum Information Processing''
(Forschungszentrum J{\"{u}}lich, 2013). All rights reserved.}

\newpage


\section{Quantum Information Processing with Atoms and Ions}

When comparing atomic systems (including ions) with solid-state systems regarding their suitability for applications in quantum information processing (QIP), one finds that atomic systems have the advantage of long coherence times for the quantum bits (qubits) because it is easy to isolate the qubits from the environment. Solid-state systems offer the advantage that only moderate effort is needed to adapt existing fabrication technologies to produce a really large number of replicas of building blocks for QIP.

The strength of each system reveals the weakness of the other. People trying to scale up atomic systems to a really large number of qubits quickly enter uncharted territory. On the other hand, many solid-state qubits suffer from short qubit coherence times which are caused by poor isolation of the qubits from their environment. The latter is due to the fact that a qubit in a solid is always surrounded by other nearby particles; and short distances tend to cause poor isolation. A recent experiment \cite{Maurer:12} achieved a qubit coherence time exceeding 1 s in a room-temperature solid. But the isolation techniques used there are not necessarily applicable to all kinds of solid-state experiments. As of now, nobody knows which systems will turn out to be most suitable for QIP in the long run.

Obviously, short coherence times can easily prevent experimentalists from performing first proof of principle experiments, whereas potential issues with future scalability will not. This is why early experiments on QIP were typically performed with photons, atom, and ions several years prior to corresponding solid-state experiments. The demonstration of quantum key distribution using photons in 1992 \cite{Bennett:92} can arguably be regarded as the first experiment on QIP. In 1995, an experiment using an atom in a cavity measured a conditional phase shift \cite{Turchette:95} and was published back-to-back with the first quantum logic gate demonstrated with ions \cite{Monroe:95}. The first solid-state quantum logic gate followed in 2003 \cite{Li:03}. Many of the insights gained from the experiments with photons, atoms, and ions are directly relevant for solid-state QIP.

Hybrid systems are an attempt to get the best of both worlds. They couple different kinds of systems; e.g.\ an early proposal suggested coupling trapped molecules with a nearby solid-state qubit \cite{Rabl:06}. The experimental implementation of such systems has started only recently. The strength of hybrid systems lies in the fact that they offer a perspective to take advantage of the long coherence times of atomic systems and of the scalability of solid-state systems.

The present chapter A3 concentrates on QIP in atoms and ions. The objective is to give a brief introduction which can clearly not be a comprehensive review. Excluded here are topics which are -- at least traditionally -- purely photonic, such as quantum cryptography \cite{Bennett:92}, dense coding \cite{Mattle:96}, quantum teleportation \cite{Boschi:98, Bouwmeester:97}, and entanglement swapping \cite{Pan:98}, because they are discussed by Barbara Kraus in chapter A5. Another field not covered here uses continuous variables (i.e.\ squeezed light and spin squeezing) instead of qubits, see e.g.\ the review article \cite{hammerer:10}. We begin with ions and then turn to atoms.

\section{Qubits in Ions}

This section briefly introduces some experimental techniques for QIP in ions. Further details on these techniques and further references can be found e.g.\ in the textbook \cite{Ghosh:95} or in the review articles \cite{Leibfried:03:RMP, Haeffner:08}. There is also some coverage on ions by Enrique Solano in chapter C2 where the focus is more on quantum simulations.

\paragraph{Cooling and Trapping.}

Ions are electrically charged and respond correspondingly strongly to electric fields. A clever design of an electrode geometry and a suitable choice of applied ac- and dc-voltages suffices to trap an ion. Once trapped, an ion can be stored for several months. The two standard geometries are called Paul trap and Penning trap. For several years now, some groups have fabricated the electrodes by micro-structuring of wires on solid-state substrates. These systems are called chip traps and offer a better perspective for scalability to a large number of qubits.

Laser light is used to cool the trapped ions. The two regimes of laser cooling employed here are called Doppler cooling and sideband cooling. The latter makes it possible to cool essentially all the population into the vibrational ground state of the trap.

\begin{figure}
\centering
\includegraphics[width=0.6\hsize]{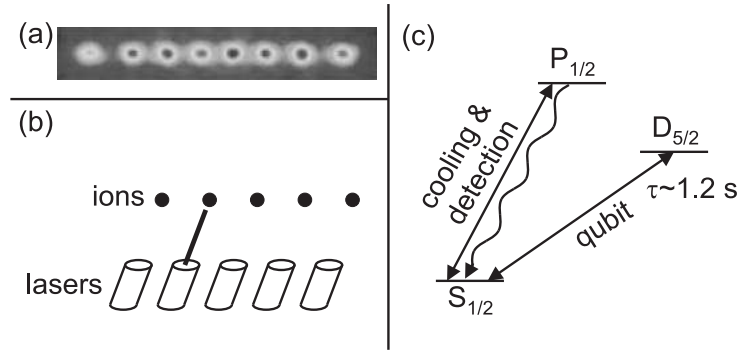}
\caption{\label{fig-ion-chain}
(a) Experimental image of a chain of 8 trapped ions. For imaging, the ions are illuminated with laser light. The florescence is recorded with a digital camera. The ions are essentially at rest. The distance between ions is $\sim$5 $\mu$m. Reproduced from Ref.\ \cite{Haffner:05}. (b) Each ion can be addressed with an individual laser beam. (c) A simplified level scheme of $^{40}$Ca$^+$. A qubit is encoded in the two-dimensional space spanned by the $S_{1/2}$ ground state and the metastable $D_{5/2}$ state with a lifetime of $\sim$1.2 s. The closed-cycling transition $S_{1/2}\leftrightarrow P_{1/2}$ is used for Doppler cooling and for state detection.
}
\end{figure}

It is possible to trap many ions simultaneously in an ion trap. If the ions are cold enough, then the Coulomb repulsion between the ions will cause the formation of an ion crystal. For an appropriate trap geometry, the ions form a linear chain in which the distance between adjacent ions is approximately 5 $\mu$m. This makes it possible to address each ion individually with a tightly focused laser beam, as illustrated in Fig.\ \ref{fig-ion-chain}(b).

\paragraph{Qubits.}

An obvious way to encode a qubit in a single ion is to pick a suitable pair of internal states. When implementing a qubit, one typically wants long coherence time combined with experimental techniques for the following general procedures: single-qubit rotations, state preparation, and state detection. For single ions one typically reaches these goals in the following way.

\paragraph{Coherence Time.}

Typical experiments encode the qubit values $|0\rangle$ and $|1\rangle$ in a pair of internal states. One strategy uses one ground state and one metastable state (optical qubit), another strategy uses two ground states with different hyperfine or Zeeman quantum numbers (hyperfine qubit or Zeeman qubit). The lifetimes of these states against spontaneous emission of a photon are extremely long. In addition to this long lifetime against population decay from the qubit states $|0\rangle$ and $|1\rangle$ ($T_1$ time), one also needs a long lifetime of the phase in a coherent superposition of the qubit states ($T_2$ time). Here, fluctuating magnetic fields often impose limitations because they introduce fluctuations in the energy difference between the qubit states.%
\footnote{If this energy difference is $\hbar\omega_{01}$, then the free evolution of the off-diagonal element of the density matrix will be given by $\rho_{01}(t)=\rho_{01}(0)e^{i\omega_{01}t}$. Hence, fluctuations of $\omega_{01}$ cause a decay of $\rho_{01}$.}
The longest $T_2$ limes will be achieved if one picks a pair of states that have the same linear Zeeman effect. A transition between such a pair of states is called clock transition because its insensitivity to magnetic-field fluctuations will also be advantageous if used for an ion clock. Another option is the use of a decoherence-free subspace, see e.g.\ Ref.\ \cite{kielpinski:01}.

\paragraph{Single-Qubit Rotations.}

In order to obtain a straightforward implementation of arbitrary sin\-gle-qu\-bit rotations, one typically chooses a pair of states that is connected by an optical transition (for an optical qubit) or by a radio-frequency transition or optical Raman transition (for a hyperfine or Zeeman qubit). To achieve high fidelity of these rotations, one needs to control the intensity, detuning, and duration of the applied pulses very accurately. In some cases, techniques based on adiabatic following, photon echoes, or geometric phases can be used to relax the control requirements on the experimental parameters somewhat.

\paragraph{State Preparation.}

Optical pumping is a standard technique for preparing the internal state of the ion. This is possible with high fidelity. If for a specific internal state no efficient optical pumping scheme is available, then one will typically use, first, optical pumping into a different state and, second, radio-frequency pulses for a coherent population transfer into the desired state.

\paragraph{State Detection.}

Many ions offer what is called a closed-cycling transition. This refers to an optical transition where an excited internal state has only one ground state into which it can decay spontaneously. If a single ion is initially in this ground state and it is excited by a laser, then this can be used to scatter a large number of photons. If the ion is in another long-lived state, however, no photon scattering occurs. Even if only a small fraction of the photons are collected and counted, one will obtain a high-fidelity internal state detection. An example is shown in Fig.\ \ref{fig-ion-chain}(c).

\section{The Cirac-Zoller Quantum Computer for Ions}

The above-listed techniques for preparation, detection, and single-qubit rotations are relatively straightforward because they involve only single-ion physics. Implementing a universal two-qubit gate, however, requires an interaction between two ions. The Coulomb repulsion between ions is an obvious candidate for this interaction. This insight lies at the heart of the seminal 1995 paper by Cirac and Zoller \cite{Cirac:95} which is one of the first proposals for a physical realization of a quantum computer.

\paragraph{Motional Sidebands.}

\begin{figure}
\centering
\includegraphics[width=0.7\hsize]{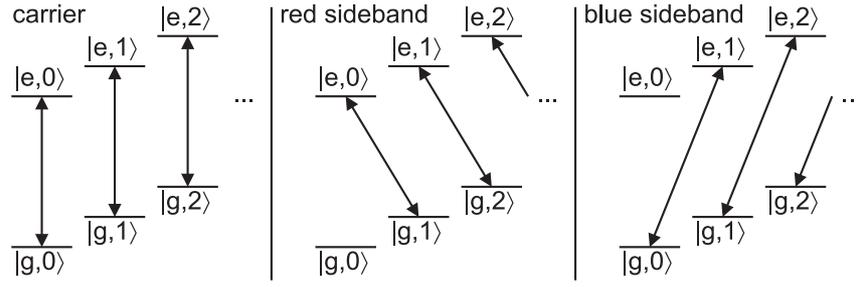}
\caption{\label{fig-sidebands}
Level scheme including motional sidebands. The levels are labeled $\ke{g,n_{CM}}$ and $\ke{e,n_{CM}}$, respectively, where $n_{CM}$ is the phonon number. The red sideband, carrier, and blue sideband drive the transitions $\lra{g,n_{CM}}{e,n_{CM}+\Delta n_{CM}}$ with $\Delta n_{CM}=-1$, $0$, and $1$, respectively. Note that the red sideband does not affect the state $\ke{g,0}$.
}
\end{figure}

For a linear chain of ions in a trap, the Coulomb interaction gives rise to normal modes of the collective motion of the cold ions. The Cirac-Zoller scheme uses the center-of-mass (CM) mode with angular frequency $\omega_{CM}$. We consider the optical transition between an internal ground state $|g\rangle$ and an excited state $|e\rangle$. This transition can change the number of phonons in the CM mode $n_{CM}$ by $\Delta n_{CM}=0$ or $\Delta n_{CM}=\pm1$. The angular frequency of the light resonant with this transition is
\begin{equation}
\omega_{eg}+\omega_{CM}\Delta n_{CM}
.
\end{equation}
In typical experiments, the width of these resonances is much less than $\omega_{CM}$, so that the motional sidebands are well resolved. A light pulse applied with this frequency will drive coherent Rabi flopping on the transitions
\begin{equation}
\lra{g,n_{CM}}{e,n_{CM}+\Delta n_{CM}}
\end{equation}
for all $n_{CM}$ simultaneously. The transitions with $\Delta n_{CM}=+1$ and $-1$ create a blue and red sideband, respectively, relative to the carrier frequency at $\omega_{eg}$. A level scheme for the $\lra{g}{e}$ transition including $n_{CM}$ is shown in Fig.\ \ref{fig-sidebands}.

$\omega_{CM}$ is typically on the order of several MHz. If a light pulse is to resolve the sideband from the carrier, then the light pulse must be longer than $1/\omega_{CM}$ to make interaction-time broadening negligible. This sets the speed limit for the Cirac-Zoller gate. In addition, other normal modes of the vibration of the ions must also be separated. This will become a serious issue if the number of ions is large, because then the sidebands of different vibrational modes lie close to each other. In passing we note that the sideband cooling mentioned earlier is based on these very sidebands.

\paragraph{The Cirac-Zoller Two-Qubit Gate.}

Using the motional sidebands, one can implement a two-qubit quantum-logic gate in a chain of $N$ ions. The $n$-th ion carries a qubit encoded in a ground state $|g_n\rangle$ and one long-lived excited state $|e_n\rangle$. There is an auxiliary excited state $|a_n\rangle$, which does not carry a qubit but will be useful during the gate operation. The gate is accomplished by consecutively applying three light pulses as follows:
\begin{enumerate}
\itemsep -1ex
\item[(i)] a red-sideband $\pi$-pulse for the $\lra{g}{e}$ transition on the $m$-th ion,
\item[(ii)] a red-sideband $2\pi$-pulse for the $\lra{g}{a}$ transition on the $n$-th ion, and
\item[(iii)] the same as (i).
\end{enumerate}
We consider the action of pulse (i). With the initial qubit values $\ke{e_m}$ or $\ke{g_m}$ combined with $n_{CM}=0$, one drives the transition
\begin{equation}
\lra{e_m,0}{g_m,1}
.
\end{equation}
The state $\ke{g_m,0}$ is unaffected by the red sideband, because the number of phonons $n_{CM}$ cannot become negative; see also Fig.\ \ref{fig-sidebands}. However, amplitude in state $\ke{e_m,0}$ is fully transferred into state $\ke{g_m,1}$, thus creating one phonon. In this process, a phase factor of $-i$ is picked up.

We turn to pulse (ii). With the initial qubit values $\ke{e_n}$ or $\ke{g_n}$ combined with $n_{CM}=0$ or 1, one drives the transition
\begin{equation}
\lra{g_n,1}{a_n,0}
.
\end{equation}
The state $\ke{g_n,0}$ is unaffected, as above. The states $\ke{e_n,n_{CM}}$ are generally not coupled at all by this light. The state $\ke{g_n,1}$, however, will undergo a $2\pi$ cycle to state $\ke{a_n,0}$ and back. The amplitude will pick up a factor of $-1$ in the process. This phase factor is crucial because it will be picked up, if and only if the initial state before pulse (i) is $\ke{e_m,g_n,0}$.

We now come to pulse (iii). As the population of all states was unchanged after pulse (ii), pulse (iii) essentially reverses the action pulse (i). It adds another phase factor of $-i$ in the transition from $\ke{g_m,1}$ to $\ke{e_m,0}$. In the absence of pulse (ii), one would simply obtain a total phase factor of $-1$ for state $\ke{e_m,0}$ from pulses (i) and (iii).

Overall, we obtain
\begin{equation}
\begin{array}{c}
\ke{g_m,g_n,0} \\ \ke{g_m,e_n,0} \\ \ke{e_m,g_n,0} \\ \ke{e_m,e_n,0} \\
\end{array}
\stackrel{\mathrm{(i)}}{\longrightarrow}
\begin{array}{r}
\ke{g_m,g_n,0} \\ \ke{g_m,e_n,0} \\ -i\ke{g_m,g_n,1} \\ -i\ke{g_m,e_n,1} \\
\end{array}
\stackrel{\mathrm{(ii)}}{\longrightarrow}
\begin{array}{r}
\ke{g_m,g_n,0} \\ \ke{g_m,e_n,0} \\ i\ke{g_m,g_n,1} \\ -i\ke{g_m,e_n,1} \\
\end{array}
\stackrel{\mathrm{(iii)}}{\longrightarrow}
\begin{array}{r}
\ke{g_m,g_n,0} \\ \ke{g_m,e_n,0} \\ \ke{e_m,g_n,0} \\ -\ke{e_m,e_n,0}
\end{array}
\begin{array}{r}
 \\  \\  \\ .
\end{array}
\end{equation}
This is a unitary transformation with matrix representation
\begin{equation}
U_\mathrm{phase}=
\left(\begin{array}{cccc}
1 & 0 & 0 & 0 \\
0 & 1 & 0 & 0 \\
0 & 0 & 1 & 0 \\
0 & 0 & 0 & -1 \\
\end{array}\right)
.
\end{equation}
This is called a conditional phase gate. This two-qubit gate is universal, i.e.\ an arbitrary computation on a quantum computer can be decomposed into two-qubit conditional phase gates and single-qubit rotations. For example, single-qubit rotations can convert this gate into a controlled NOT (CNOT) gate with matrix representation
\begin{equation}
\label{U-CNOT}
U_\mathrm{CNOT}=
\left(\begin{array}{cccc}
1 & 0 & 0 & 0 \\
0 & 1 & 0 & 0 \\
0 & 0 & 0 & 1 \\
0 & 0 & 1 & 0 \\
\end{array}\right)
\end{equation}
which is also universal.

\begin{figure}
\centering
\includegraphics[width=0.4\hsize]{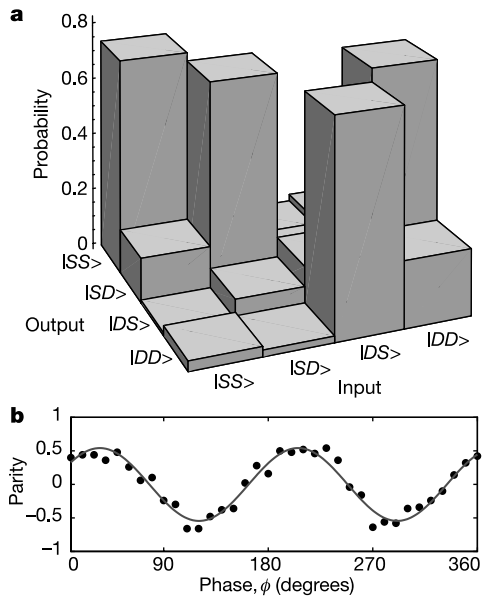}
\caption{\label{fig-Cirac-Zoller-data}
Experimental data on the Cirac-Zoller gate. (a) Truth table for the gate, showing the measured modulus squared of the matrix elements of $U_\mathrm{CNOT}$ from Eq.\ \eqref{U-CNOT}. (b) Demonstration of phase coherence. Reproduced from Ref.\ \cite{Schmidt-Kaler:03}.
}
\end{figure}

Shortly after publication of the proposal, a first experiment \cite{Monroe:95} demonstrated a version of the Cirac-Zoller gate that uses two degrees of freedom of a single ion as the qubits.  The full Cirac-Zoller gate for two ions was experimentally demonstrated in 2003 \cite{Schmidt-Kaler:03}. Data from this experiment are shown in Fig.\ \ref{fig-Cirac-Zoller-data}. Part (a) shows the measured modulus squared of all matrix elements of $U_\mathrm{CNOT}$ from Eq.\ \eqref{U-CNOT}. This is called truth table. In principle, the phases of all 16 matrix elements remain to be measured. In practice, however, one often performs a test of phase coherence in an appropriate basis, such as the one shown in Fig.\ \ref{fig-Cirac-Zoller-data}(b), and considers this as sufficient to show that the gate operates properly.

\paragraph{Generalizations.}

The generalization of this scheme to an $n$-qubit conditional phase gate for arbitrary $n$ is straightforward \cite{Cirac:95}. Having such gates is desirable, because if a complicated algorithm is decomposed into $n$-qubit gates, then the number of required operations will decrease with increasing $n$.

The Cirac-Zoller gate heavily relies on the assumption that $n_{CM}=0$ initially. In an experiment, the probability for $n_{CM}=1$ can be made small, but never exactly zero. This corresponds to a nonzero temperature in the population of this mode. This limits the achievable gate fidelity. A solution lies in the M{\o}lmer-S{\o}rensen gate, proposed in Refs.\ \cite{Molmer:99, Sorensen:99, Sorensen:00}, which can tolerate nonzero temperatures much better. An addition advantage is that this gate does not require individual addressing of the ions during the gate operation.

Any universal two-qubit gate must be able to generate a maximally entangled state from an appropriately chosen, separable input state. This is called entangling gate operation. It is frequently used when experimentally demonstrating that a logic gate works as advertised. For example, the first experimental demonstration of the M{\o}lmer-S{\o}rensen gate entangled up to 4 ions \cite{Sackett:00}. More recent experiments entangled up to 14 ions \cite{Haffner:05, Monz:11}. For entangled states of so many particles, it is a serious challenge to conceive and carry out a measurement procedure that proves the entanglement.

Another two-ion gate is the geometric phase gate, proposed in Ref.\ \cite{Milburn:00}. It shares the advantages of the M{\o}lmer-S{\o}rensen gate, but is less sensitive to various experimental imperfections. Its first experimental demonstration in 2003 \cite{Leibfried:03:Nature} already achieved a fidelity of 97\%.

\section{Qubits in Atoms}

A related line of research pursues QIP in neutral atoms. Coherence time, state preparation, state detection, and single-qubit rotations are largely similar to ions. Again, light forces created with lasers are the standard tool for cooling the particles. But now, they also serve for trapping the particles, see the textbook \cite{metcalf:99} for details and for further references.

\paragraph{Cooling and Trapping.}

The magneto-optical trap (MOT) is an extension of the Doppler cooling scheme. It is used for capturing up to $10^{10}$ atoms from an atomic beam or from a vapor and for cooling the atoms to temperatures of $\sim$100 $\mu$K. After collection and cooling in a MOT, many experiments use what physicist call an optical dipole trap and biologists call optical tweezers. This trap is made of a focused laser beam that is far red detuned from all atomic transitions. The light induces an oscillating electric dipole moment in the atom, which experiences a potential energy in the light field. The atoms are attracted to regions of high light intensity. Optical dipole traps can be several mK deep, but many are much shallower.

\paragraph{Van-der-Waals Interaction.}

As discussed above, an interparticle interaction is needed to build a quantum logic gate. In this respect it is problematic, that the interaction between neutral ground-state atoms is much weaker than between ions. Instead of a charge, the interaction relies on the electric polarizability of the atoms. At large interatomic distance $r$, this creates a van-der-Waals tail of the atom-atom interaction potential, described by $V(r)=C_6/r^{6}$. The $C_6$ coefficient is typically on the order of $10^3$ atomic units (1 a.u$.=9.57\times 10^{-80}$ Jm$^6$). The weakness of this interaction makes it difficult to achieve gate-operation times that are short enough to be interesting. For example, to achieve a gate operation time of 1 ms, one would need at least $V=2\pi\hbar \times 1$ kHz. This requires an interparticle separation of $r\sim20$ nm, which makes individual addressing with laser light impractical. For comparison, the Coulomb interaction between two ions at a distance of 5 $\mu$m amounts to $\sim 2\pi\hbar\times 70$ GHz.

\paragraph{Rydberg States.}

There are various options how to solve the problem of the weakness of the interaction. One option is the use of Rydberg states. They have a much larger polarizability and have recently been used to realize a universal quantum logic gate for neutral atoms \cite{Wilk:10, Isenhower:10} and to build a single-photon source \cite{Dudin:12}.

\paragraph{Atoms and Photons.}

Another option to influence one atom with another is by emitting a single photon from the one atom and absorbing it in the other. The polarization of the photon can carry a qubit. The tricky part is to emit only one photon and to make the emission directed and the absorption efficient. A brief look at the absorption process already illustrates the problem and hints at possible solutions. An atom has an absorption cross section of $\sigma=3\lambda^2/2\pi$ for light at wavelength $\lambda$, resonant with a closed-cycling transition. It is very difficult to focus a laser beam down to small enough a spot size to obtain deterministic absorption in a single atom. However, there are two strategies to effectively enhance the atom-photon interaction.

\paragraph{Atomic Ensembles.}

One strategy is to use an ensemble, consisting of a large number of atoms. The single photon can be absorbed by any atom in the ensemble. It remains unknown, which atom absorbed the photon. Hence, the absorption creates a collective excitation involving all atoms. The collective excitation can carry a qubit. The absorption cross section is effectively multiplied by the number of atoms. The absorption probability can reach 100\%, at least in principle, and atomic ensembles can thus be used to realize QIP protocols.

\paragraph{Atoms-Cavity Systems.}

The other strategy is to place a single atom inside an optical resonator with highly reflective mirrors. If a photon is coupled into a resonant cavity mode, then it will bounce back and forth between the mirrors many times. To lowest approximation, one can think of this as the photon traveling through the atom many times, thus enhancing the absorption probability by the number of bounces from the mirrors.

\bigskip

QIP experiments based on either strategy, ensembles or cavities, are described in the following sections. We start our description with the emission in Sec.\ \ref{sec-source}. Combining it with the absorption makes it possible to build quantum memories, as described in Sec.\ \ref{sec-memory}. Based on these basics, more complex tasks can be mastered. As an example, we consider an experiment that generates remote entanglement in Sec.\ \ref{sec-remote}. The techniques presented here offer perspectives for building quantum gates \cite{Pellizzari:95, duan:04, Rispe:11, Vo:1211.7240} and quantum networks \cite{cirac:97, kimble:08, Ritter:12}.

\section{Atoms as Single-Photon Sources}
\label{sec-source}

We consider the process of photon emission. One needs a source for the triggered generation of a single photon with a collimated transverse mode. This can be realized either using a single atom in a cavity or an atomic ensemble.

\begin{figure}
\centering
\includegraphics[width=0.8\hsize]{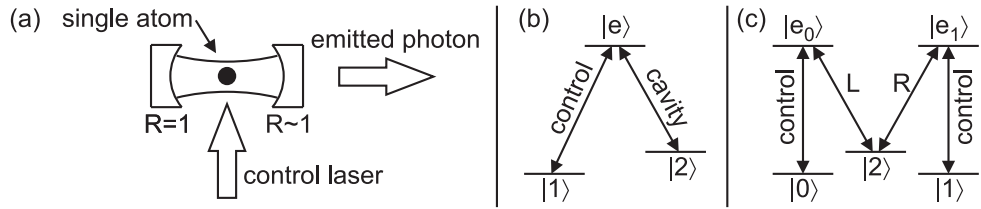}
\caption{\label{fig-photon-pistol}
Deterministic single-photon source using a single atom. (a) A single atom is excited by a control laser. The scattered photon is preferentially emitted into the cavity mode, from which it is emitted due to the nonzero transmission of one mirror. (b) Simplified atomic level scheme. (c) Extended level scheme. An atomic qubit, encoded in states $\ke{0}$ and $\ke{1}$, can be mapped onto the polarization qubit of the single photon.
}
\end{figure}

\paragraph{Atom-Cavity Systems.}

A single atom in a Fabry-Perot resonator is well suited for generating a single photon, as illustrated in Fig.\ \ref{fig-photon-pistol}. The internal states of the atom form a $\Lambda$ scheme, with two ground states, labeled $\ke{1}$ and $\ke{2}$, each coupled to the joint excited state $\ke{e}$. One mode of the resonator is tuned into resonance with the transition $\lra{2}{e}$. To drive the process, a control laser beam containing many photons is sent onto the atom from the side. This light resonantly couples the atom in the initially prepared state $\ke{1}$ to the excited state $\ke{e}$. From here, the atom will decay back into a ground state. This can either be an undesired, spontaneous process with photon emission essentially into the full solid angle or it can be an emission into the resonator mode that is resonant with the $\lra{2}{e}$ transition. Using high-reflectivity mirrors, small mode volume, and adiabatic techniques \cite{kuhn:02} (see also appendix \ref{sec-3-level}) one can make the emission into the resonator mode the dominant process \cite{Ritter:12}. One deliberately makes one mirror less reflective than the other. Hence, the photon almost always leaves the resonator through this mirror.

All properties of the emitted photon are well controlled: a useful, collimated transverse mode, the polarization, the center frequency, the emission time, and the spatiotemporal wave packet. The temporal shape of the control laser pulse can be used to tailor the spatiotemporal wave packet of the single photon. As only one atom is inside the resonator and as the final atomic state is not coupled to the control laser, only one photon can be emitted. The efficiency for emitting the photon into the cavity mode is $\sim$60\% in present experiments \cite{Ritter:12} and it can theoretically reach 100\%. One can easily extend the atomic level scheme to map an atomic qubit encoded in states $\ke{0}$ and $\ke{1}$ onto the polarization qubit of the single photon; see Fig.\ \ref{fig-photon-pistol}(c). $L$ and $R$ denote the two circular photon polarizations.

The mathematical description of the flying single photon generated here is based on its spatiotemporal mode function $u(z,t)=u(0,t-z/c)$ with time $t$, longitudinal coordinate $z$, and light speed $c$. A Fourier decomposition of this mode function into plane waves $e^{-i\omega(t-z/c)}$ with angular frequencies $\omega$ combined with the standard quantization of these plane waves yields creation and annihilation operators $\hat a^\dag$ and $\hat a$ for this mode. The single-photon state for this mode is obtained by applying this creation operator to the vacuum.

A Fock state with exactly one photon is non-classical. The non-classicality can be quantified using the pair correlation function \cite{Glauber:63}
\begin{equation}
g^{(2)}
= \frac{\lr{\hat a^\dag \hat a^\dag \hat a \hat a}}{\lr{\hat a^\dag \hat a}^2}
= \frac{\sum_n n(n-1)p_n}{(\sum_n np_n)^2}
\end{equation}
where $p_n$ denotes the probability for detecting $n$ photons. A light pulse emitted by a laser has a Poisson distribution of the $p_n$, yielding $g^{(2)}=1$. States with $g^{(2)}<1$ are non-classical. They are called anti-bunched. Single-photon sources typically operate in the regime $p_{n+1}\ll p_n$ for $n\neq0$ so that
\begin{equation}
\label{g2-sim}
g^{(2)}
\sim \frac{2p_2}{p_1^2}
\end{equation}
characterizes the probability $p_2$ for the undesired production of two photons normalized to the classical limit. A recent atom-cavity experiment reported a value as low as $g^{(2)}=0.01$ \cite{lettner:11}.

\paragraph{Atomic Ensembles and Heralding: The DLCZ Single-Photon Source.}

An atomic ensemble can also serve as a single-photon source using a heralding scheme which is a downsized version of the Duan-Lukin-Cirac-Zoller (DLCZ) proposal \cite{Duan:01, Duan:02} for a quantum repeater \cite{briegel:98, Sangouard:11}. The DLCZ single-photon source was experimentally demonstrated in Ref.\ \cite{Kuzmich:03}.

The basic idea is the following. If one had a detector that measured the photon number $n$ nondestructively, then this device could herald (i.e.\ announce) $n$ and one could discard all events with $n\neq1$ to obtain a perfect single-photon source.

However, typical photon-counting detectors (avalanche photodiodes, photomultiplier tubes etc.) have two properties that make an immediate application of this idea unfeasible. First, they have a dead time after each detection event during which they cannot register any further photons and, second, they destroy the photons. We will discuss now how one can nevertheless construct a useful heralding scheme using such detectors.

First, we take the dead time into account but assume that the detector would not destroy the photons. The dead time implies that a detection event indicates that a nonzero number of photons arrived. Such a detector will still be useful if one starts from a Poissonian photon source with mean photon number per pulse much below 1. Such a source already has the desired property $p_{n+1}\ll p_n$ for $n\neq0$. The strategy is to let the detector herald the pulses with $n\neq0$ and to discard the other pulses. One can easily estimate $g^{(2)}$ for such a source. We denote the (conditional) probabilities for the heralded events as $\widetilde p_n$ and the corresponding pair-correlation function as $\widetilde g^{(2)}$. Ideally, we obtain $\widetilde p_0=0$ and $\widetilde p_n= p_n/(1-p_0) \sim p_n/p_1$ for $n\neq0$. Eq.\ \eqref{g2-sim} yields
\begin{equation}
\label{g2-DLCZ}
\widetilde g^{(2)}
\sim \frac{2\widetilde p_2}{\widetilde p_1^2}
\sim \frac{2p_2}{p_1}
\sim p_1 g^{(2)}
= p_1
\ll 1
\end{equation}
where we used $g^{(2)}=1$ for the unheralded Poisson distribution. The smaller $p_1$, the better the suppression of $\widetilde g^{(2)}$. This scheme would yield an anti-bunched single-photon source if only we had such a detector.

\begin{figure}
\centering
\includegraphics[width=0.8\hsize]{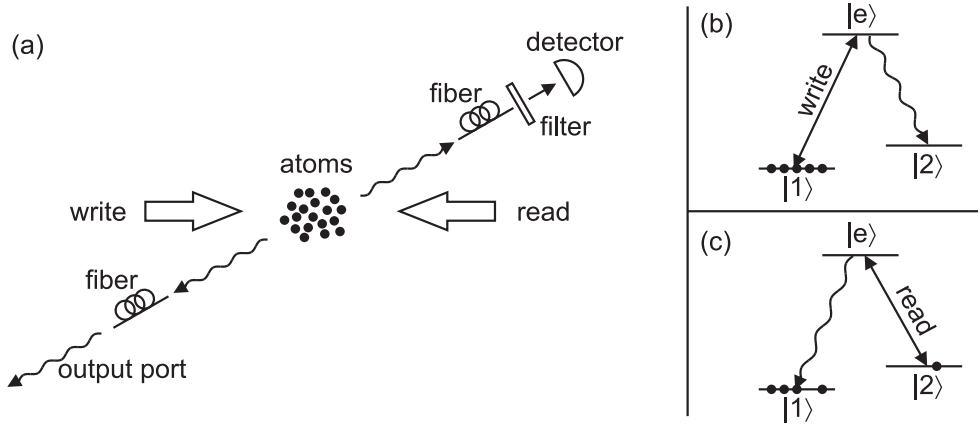}
\caption{\label{fig-DLCZ}
DLCZ source for single photons. (a) An atomic ensemble is illuminated by a weak light pulse for writing. If the detector registers a scattered photon, then this will project the ensemble into a state with one collective excitation. A subsequent light pulse for reading converts this into a single photon propagating towards the output port. (b) and (c) simplified atomic level schemes for the write and read process, respectively. The circles symbolize the atom number in each state before application of the pulse.
}
\end{figure}

We now turn to the second issue: the destruction of the photons in the detector. This is where atomic ensembles can help. Again, one uses internal atomic states that form a $\Lambda$ scheme, as shown in Fig.\ \ref{fig-DLCZ}. Using optical pumping, all atoms are initially prepared in the ground state $\ke{1}$. A pulse of laser light, called write pulse, illuminates the ensemble. The write pulse is resonant with the $\lra{1}{e}$ transition and it must be weak (in a sense to be discussed below). The detector in Fig.\ \ref{fig-DLCZ} can detect scattered photons. In front of the detector, there is an optical fiber for transverse mode selection and a frequency filter (e.g.\ a filter cavity) that transmits only light resonant with the $\lra{2}{e}$ transition. If the detector clicks, i.e.\ registers a photon, then this measurement result will be used as a herald. This event projects the ensemble into a state where an atom has been transferred to state $\ke{2}$. A subsequent read pulse with many photons, resonant with the $\lra{2}{e}$ transition, optically pumps this atom back into the original state $\ke{1}$ under emission of a photon resonant with the $\lra{1}{e}$ transition.

We now consider $g^{(2)}$. Each photon scattered from the write beam into the detector will ideally leave behind exactly one atom in state $\ke{2}$ so that the subsequent read process will ideally emit exactly one photon. This part is deterministic. However, a Poisson distribution does appear in this scheme; namely for the number of photons scattered from the write beam into the detector. Experimental parameters must be chosen such that the mean number of these photons is much below 1. In other words, one needs a heralding probability $1-p_0 \ll1$. Hence, Eq.\ \eqref{g2-DLCZ} is applicable and the readout yields an anti-bunched single-photon source.

We now study the emission direction of the final photon. When the detector clicks, it is known that exactly one atom was transferred into state $\ke{2}$. The probability that two photons reached the detector is negligible because $1-p_0\ll1$. However, it remains unknown which atom was transferred. Hence, after the click the ensemble is in a coherent superposition of all these possibilities%
\footnote{The state vector for each atom lies in the tensor product space of spatial part and spin part. Eqs.\ \eqref{magnon} and \eqref{final} use the position representation for the spatial part but no representation for the spin part. The total state vector $\ke{\psi}$ for all atoms is obtained by applying $\int d^3x_1 \ke{x_1}\cdots \int d^3x_N \ke{x_N}$ from the left to Eqs.\ \eqref{magnon} or \eqref{final}.
}
\begin{equation}
\label{magnon}
\frac1{\sqrt N} \sum_{n=1}^N e^{i (\bm k_w-\bm k_d)\cdot\bm x_n}
\psi_n(x_n) \ke{2}_n
\prod_{{\scriptstyle k=1 \atop \scriptstyle k\neq n}}^{N} \psi_k(x_k) \ke{1}_k
.
\end{equation}
Here, $\ke{1}_n$ and $\ke{2}_n$ denote the internal states of the $n$-th atom. $\psi_n(x_n)$ denotes the initial spatial wave function of the $n$-th atom. We assumed that all $N$ atoms were initially in internal state $\ke{1}$ and in a spatial product state. If the $n$-th atom is transferred into state $\ke{2}$, then its spatial wave function will pick up a phase factor $e^{i (\bm k_w-\bm k_d)\cdot\bm x_n}$ due to the differential photon recoil, where $\bm k_w$ and $\bm k_d$ are the wave vectors of the write photon and the detected photon. Typically, states $\ke{1}$ and $\ke{2}$ are two ground states with different hyperfine quantum numbers. Hence, the state in Eq.\ \eqref{magnon} describes a spin-wave excitation. Moreover, the fact that the number of transferred atoms is exactly one means that this state represents a single magnon, the quasi-particle of the spin wave.

When the read pulse is applied, the $n$-th atom is transferred back into state $\ke{1}$ and its spatial wave function picks up a phase factor $e^{i (\bm k_r-\bm k_f)\cdot\bm x_n}$, where $\bm k_r$ and $\bm k_f$ are the wave vectors of the read photon and the final photon. The final state of the ensemble is
\begin{equation}
\label{final}
\frac1{\sqrt N} \left( \sum_{n=1}^N e^{i (\bm k_w-\bm k_d+\bm k_r-\bm k_f)\cdot\bm x_n} \right)
\prod_{k=1}^{N} \psi_k(x_k) \ke{1}_k
.
\end{equation}
In principle, $\bm k_f$ can have any direction. However, if the condition
\begin{equation}
\bm k_w-\bm k_d + \bm k_r -\bm k_f = 0
\end{equation}
is met, then all phase factors in Eq.\ \eqref{final} will be unity and all terms in the sum will interfere constructively. This gives the probability for this emission direction a collective enhancement factor of $N$, the number of atoms \cite{Duan:02}. For large $N$, the emission into other directions becomes negligible. In Fig.\ \ref{fig-DLCZ}, collective, directed emission towards the output port is obtained. The generated photon has a well-defined, collimated transverse mode.%
\footnote{The directed character of the emission also implies that scattering of photons from the write beam into directions other than towards the detector is not a concern, because upon readout this will generate photons not sent towards the output port. Hence, the value of $g^{(2)}$ measured in the direction of the output port does not deteriorate.
}
Here we assumed that the radius of the ensemble is much larger than the optical wavelength. Otherwise, the phase factors do not change noticeably within the radius of the ensemble and the generated single photon will look as if it had been diffracted from an aperture with the size of the ensemble.

All properties of the emitted photon are well controlled: a useful, collimated transverse mode, the polarization, the center frequency, the emission time, and the spatiotemporal wave packet. The temporal shape of the read pulse can be used to tailor the spatiotemporal wave packet of the single photon. Moreover, it is possible to extend the atomic level scheme to map an atomic qubit onto the polarization qubit of the photon \cite{matsukevich:06, tanji:09}.

Ref.\ \cite{Laurat:06} reports $\widetilde g^{(2)}=0.007$ for a heralding probability of $p_1 \sim 4\times 10^{-5}$ and a photon-emission efficiency of $\sim$45\% conditioned on heralding. Here, $\widetilde g^{(2)}$ is a factor of $\sim$200 worse than expected from the simple estimate Eq.\ \eqref{g2-DLCZ}. Several technical imperfections contribute to this deviation, such as non-unit detector efficiencies, detector dark counts, imperfect transverse mode matching etc. Note that in this experiment, the number of photons per write pulse was $\sim10^4$. This number does not need to be small compared to one.

The disadvantage of a DLCZ source is that the probability $p_1$ that a herald is obtained must be made tiny if one wants a good suppression of $\widetilde g^{(2)}$. This makes count rates low and experiments may become tiresome. Unlike an atom-cavity system, the DLCZ source is inherently probabilistic. On the other hand, atom-cavity systems have the drawback that they are much harder to build.

\section{Atoms as Quantum Memories for Photonic Qubits}
\label{sec-memory}

In addition to the controlled emission of a single photon, one also needs a mechanism for the controlled absorption of a single photon. If one wanted to characterize only the absorption process, then one would need a direct measurement of the atomic state after the absorption. This is often difficult. Instead, one often maps the atomic state back to another photon in a subsequent emission process. If the photon carries a qubit, this combination of absorption and emission will represent a quantum memory \cite{lvovsky:09}.

In addition to the ability to store the qubit values $\ke{0}$ and $\ke{1}$, a quantum memory must also be able to store any coherent superposition thereof. This task cannot be performed by a classical apparatus, which performs a measurement, stores the result in a classical memory, and finally rebuilds the state from scratch upon readout. If this were possible, then the information in the classical memory could be copied many times and the apparatus could emit an arbitrary number of replicas of the single-photon qubit, thus violating the no-cloning theorem.%
\footnote{If a von-Neumann measurement is performed, one will be forced to choose one basis and will obtain only one measurement result because there is only one incoming photon. It is obvious that this cannot yield all the information about the qubit state. However, if a series of weak measurements is performed, the situation will get more complex, but the argument with the no-cloning theorem remains valid.
}
Hence, the qubit must be encoded in the state of a quantum system during the whole storage time.

The performance of a quantum memory is characterized by several figures of merit:
\begin{itemize}
\itemsep -1ex
\item The write-read efficiency $\eta$ is the ratio of the retrieved photon number over the incoming photon number.
\item The fidelity $F=\operatorname{Tr}(\rho_\mathrm{in}\rho_\mathrm{out})$ describes how well the density matrix of the incoming qubit state $\rho_\mathrm{in}$ is reproduced in the output state $\rho_\mathrm{out}$. While some input states might be well preserved, others might not. To express this in one figure of merit, one averages $F$ over the full Poincar\'{e} sphere of pure photon states, obtaining the average fidelity $\lr F$. The classical limit is $\lr F \leq 2/3$. An ideal quantum memory yields $\lr F=1$.
\item The lifetime of the efficiency characterizes how quickly $\eta$ decays over time.
\item The coherence time characterizes how quickly $\lr F$ decays over time.
\end{itemize}

\paragraph{Atom-Cavity Systems.}

The single-photon generation process for a single atom in a cavity, discussed in Sec.\ \ref{sec-source}, can be reversed in time, resulting in controlled absorption, as proposed in Ref.\ \cite{cirac:97}. The atom is initially prepared in state $\ke{2}$ of Fig.\ \ref{fig-photon-pistol}(b). For an almost arbitrary spatiotemporal wave packet of the incoming photon, one can tailor the temporal shape of the control pulse such that the photon is absorbed in the single atom with high probability, with a theoretical limit of 100\% \cite{Dilley:12}.

A recent experiment \cite{Ritter:12} used an atom-cavity system as a quantum memory and reported a write-read efficiency of $\eta =10$\% and an average fidelity of $\lr F= 92$\%. The coherence time in a similar experiment \cite{Specht:11} was $\sim$0.2 ms.

\paragraph{Atomic Ensembles: Slow Light and EIT.}

Light storage is also possible in an atomic ensemble. It is essentially the time-reversed version of the readout part of the DLCZ single-photon source. More specifically, one uses the techniques of slow and stopped light in the context of electromagnetically induced transparency (EIT) \cite{fleischhauer:05}.

\begin{figure}
\centering
\includegraphics[width=0.18\hsize]{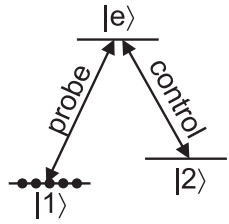}
\caption{\label{fig-EIT-levels}
Level scheme for EIT. The circles symbolize the atom number in each state before application of the probe pulse.
}
\end{figure}

Consider the $\Lambda$ system of Fig.\ \ref{fig-EIT-levels} with population initially in state $\ke{1}$ and with applied control light resonant with the $\lra{2}{e}$ transition. Let probe light (the incoming single photon) resonant with the $\lra{1}{e}$ transition impinge on the atomic ensemble. The control light drastically modifies the refractive index $n=\sqrt{1+\chi}\sim 1+\frac12\chi$ for the probe light. Here, $\chi$ is the electric susceptibility. Its real and imaginary parts are related to dispersion and absorption, respectively.

\begin{figure}[!b]
\centering
\includegraphics[width=0.9\hsize]{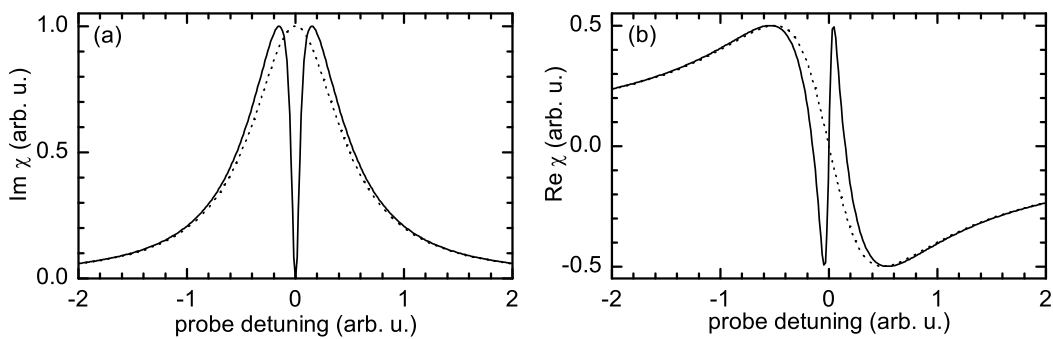}
\caption{\label{fig-EIT-chi}
EIT and slow light. (a) Imaginary and (b) real parts of the electric susceptibility $\chi$ are shown. The dotted lines (without control light) show a Lorentzian absorption profile for $\Im\chi$ and a dispersive profile for $\Re\chi$. The solid lines (with control light) show a narrow transparency window for $\Im\chi$ and a corresponding slow-light region for $\Re\chi$.}
\end{figure}

They are shown in Fig.\ \ref{fig-EIT-chi}. The Lorentzian absorption profile in $\Im\chi$ adopts a narrow transparency window due to the control light. This transmission phenomenon is called EIT; see appendix \ref{sec-3-level} for understanding its origin. In line with the Kramers-Kronig relations, a feature of similar width is obtained in the dispersive profile in $\Re\chi$. It has a steep, linear slope. The group velocity of a probe light pulse is inversely proportional to the steepness of this slope. The lower the control intensity, the narrower the EIT window and the smaller the group velocity, for reasons discussed in appendix \ref{sec-3-level}. A slowdown of the group velocity of light by 7 orders of magnitude was demonstrated in Ref.\ \cite{hau:99}.

The slowdown of the group velocity causes a drastic spatial compression of the pulse. If a probe pulse is 300 ns long, it will have a longitudinal extension of $\sim$100 m in vacuum. 7 orders of magnitude slowdown will compressed it to $\sim$10 $\mu$m which is small enough to fit into the atomic ensemble. Once the pulse is completely inside the ensemble, one ramps the control intensity to zero. This reduces the  group velocity to zero. This is called stopped light. The incoming probe light was absorbed in a controlled fashion. This absorption process is the time reversed version of the readout part of the DLCZ source. Each probe photon is converted into a single excitation in the initially empty state $\ke{2}$. For a single probe photon, the state after storage looks much like Eq.\ \eqref{magnon}. Again, the relative phases in the coherent superposition store the information about the propagation direction of the probe light. At a later time, the control light is ramped back on. This corresponds to the readout part of the DLCZ source (without time reversal). It reestablishes the probe pulse. Unlike the DLCZ source, the EIT memory is not inherently probabilistic.

Note that the slow-light effects will only be obtained if states $\ke{2}$ and $\ke{e}$ carry only a small fraction of the total atomic population at all times. Moreover, to obtain high write-read efficiency, it must be possible to longitudinally compress the complete pulse into the ensemble and at the same time have an EIT transparency window that is wide enough in frequency space to avoid irreversible absorption. This condition can only be met if the optical depth $\sigma \varrho L$ is much larger than unity, where $L$ is the length of the ensemble, $\varrho$ the number of atoms per volume, and $\sigma$ the absorption cross section that one atom represents for resonant probe light.

EIT-based storage and retrieval was demonstrated in 2001 for classical probe light pulses \cite{liu:01, phillips:01, turukhin:01} and in 2005 for anti-bunched single photons \cite{chaneliere:05, eisaman:05}. A lifetime of the efficiency of half a second was reported in Ref.\ \cite{zhang:09}. A recent experiment \cite{Riedl:12} realized a quantum memory for a photonic qubit in a Bose-condensed atomic ensemble and achieved $\eta=0.53$, $\lr F=1.000\pm0.004$, a lifetime of the efficiency of 0.5~ms, and a coherence time of 1.1 ms.

\section{Creation of Remote Entanglement}
\label{sec-remote}

As an example for the variety of possibilities offered by the techniques described above, we note that entanglement between two atomic systems at remote locations has been created in several experiments \cite{Ritter:12, matsukevich:06, julsgaard:01, chou:05, chou:07, moehring:07, yuan:08}. Specifically, we consider a recent experiment \cite{lettner:11} that created remote entanglement between a single atom in a cavity and an atomic Bose-Einstein condensate (BEC).

\begin{figure}
\centering
\includegraphics[width=\hsize]{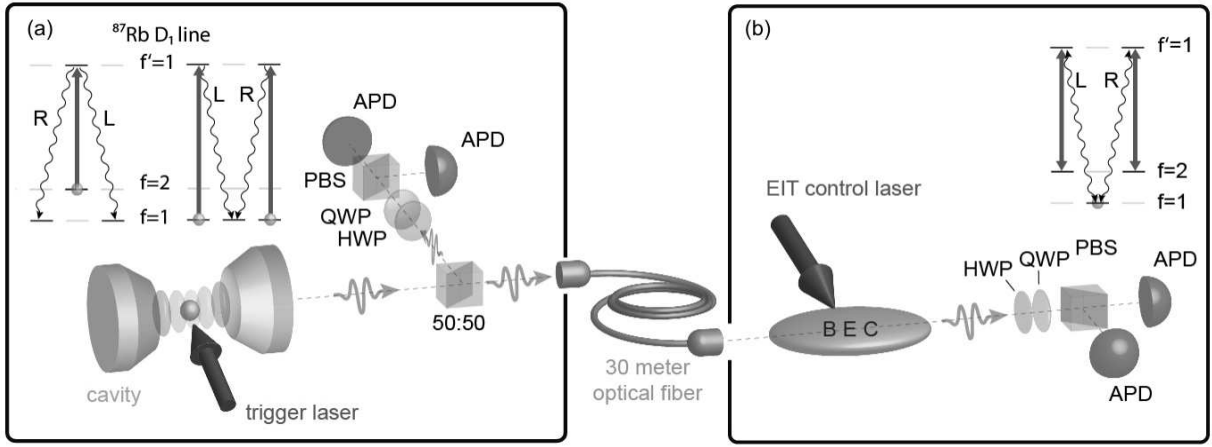}
\caption{\label{fig-entanglement}
Scheme of an experiment for creating remote entanglement. (a) An atom-cavity system serves as a single-photon source and creates atom-photon entanglement upon photon emission. (b) A BEC serves as a quantum memory for the photon. Storage in the BEC maps the photonic qubit onto the BEC, thus creating atom-BEC entanglement. At a later time, both qubits are converted into photons, which are measured. Reproduced from Ref.\ \cite{lettner:11}.
}
\end{figure}

The single atom serves not only as a single-photon source, but also as a source of two-particle entanglement \cite{wilk:07}. This is achieved with the level scheme shown in Fig.\ \ref{fig-entanglement}(a). A light pulse from the trigger laser causes the emission of a single photon as described in Sec.\ \ref{sec-source}. However, the photon can be $L$ or $R$ polarized. The final atomic state depends on the photon polarization. It is unknown which of the two paths was taken, so that the system is in a coherent superposition of the two possibilities. Hence after the emission, the Zeeman state of the single atom is entangled with the polarization of the single photon
\begin{equation}
\ke{\psi_\mathrm{atom\otimes photon}}
= \frac1{\sqrt2} \left( \ke{1,1}\otimes \ke{L} - \ke{1,-1}\otimes \ke{R} \right)
.
\end{equation}
Here, $\ke{f,m_f}$ denotes the hyperfine and Zeeman quantum state of the atom. This single photon is transported through an optical fiber to a BEC in another laboratory. Here, the photon is absorbed using EIT, as described in Sec.\ \ref{sec-memory}. The BEC serves as a quantum memory. The single photon is stored as a single magnon. The polarization qubit is stored in a Zeeman qubit using the extended internal level scheme of Fig.\ \ref{fig-entanglement}(b). $\ke{f,m_f}$ denote the hyperfine and Zeeman quantum state of the single magnon. The storage establishes entanglement between the single atom and the BEC
\begin{equation}
\ke{\psi_\mathrm{atom\otimes BEC}}
= \frac1{\sqrt2} \left( \ke{1,1}\otimes \ke{2,-1} - \ke{1,-1}\otimes \ke{2,1} \right)
.
\end{equation}
At a later time, the memory is read out, mapping the magnon qubit back to the polarization of the emitted photon. In addition, a second trigger pulse sent onto the single atom maps the Zeeman qubit of the single atom onto a second photon. After both emission processes, the two photons are entangled
\begin{equation}
\label{Bell}
\ke{\psi_\mathrm{photon\otimes photon}}
= \frac1{\sqrt2} \left( \ke{R}\otimes \ke{L} - \ke{L}\otimes \ke{R} \right)
.
\end{equation}
The polarization of each photon is measured with avalanche photodiodes (APD) placed behind a polarizing beam splitter (PBS). A combination of a quarter wave plate (QWP) and a half wave plate (HWP) in front of each PBS makes it possible to select an arbitrary measurement basis on the two Poincar\'{e} spheres.

The fidelity of the measured final state with the desired, maximally-entangled state of Eq.\ \eqref{Bell} is $F=0.95$. This characterizes the concatenation of all four processes: photon generation, storage, retrieval, and second photon generation. The coherence time for the decay of $F$ is 0.1 ms.

\section*{Appendix}

\appendix

\section{Lambda Systems: STIRAP and Dark-State Polaritons}
\label{sec-3-level}

\begin{figure}
\centering
\includegraphics[width=0.55\hsize]{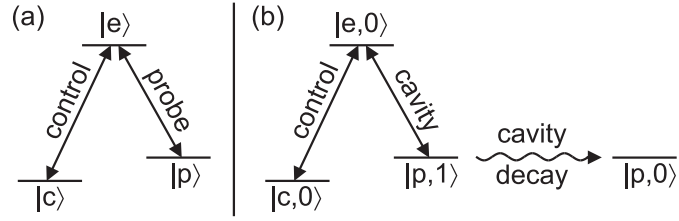}
\caption{\label{fig-vSTIRAP}
(a) Generic $\Lambda$ system. (b) Vacuum STIRAP. The atomic level scheme of Fig.\ \ref{fig-photon-pistol}(b) is extended by including the photon number, $n$, in the cavity mode. The states are labeled $\ke{c,n}$, $\ke{e,n}$, and $\ke{p,n}$. Transmission of a photon through a cavity mirror is labeled cavity decay.}
\end{figure}

This appendix offers additional insights into $\Lambda$ systems. This is relevant for the suppression of spontaneous emission in the atom-cavity single-photon source in Sec.\ \ref{sec-source} and for the emergence of EIT and slow light in Sec.\ \ref{sec-memory}. A generic $\Lambda$ system is shown in Fig.\ \ref{fig-vSTIRAP}(a).

\paragraph{Two-Level Atom.}

Consider a single atom. Control light couples the atomic states $\ke{c}$ and $\ke{e}$ in Fig.\ \ref{fig-vSTIRAP}(a). Let the electric field of the control light be $\bm E_c(t)=\Re(\bm E_{c,0} e^{-i\omega_c t})$ with amplitude $\bm E_{c,0}$ and angular frequency $\omega_c$. Let $\hbar\omega_{ce}$ denote the energy difference of states $\ke{c}$ and $\ke{e}$. Assuming that the atom is much smaller than the optical wavelength, one can describe the atom-light interaction in the electric-dipole approximation $V(t)=-\bm \mu \cdot \bm E_c(t)$, where $\bm \mu$ is the operator of the electric-dipole moment. The control transition has a matrix element $\bm \mu_{ec} =\lr{e|\bm \mu|c}$. The operator $\bm \mu$ has negative parity and the states $\ke{c}$ and $\ke{e}$ have well-defined parity. Hence, the diagonal matrix elements $\lr{c|\bm \mu|c}$ and $\lr{e|\bm \mu|e}$ vanish. The Hamiltonian for the two-level system with basis $(\ke{c},\ke{e})$ reads $H=\hbar \omega_{ce} \kb{e}{e} - (\bm \mu_{ec} \cdot \bm E_c(t)\kb{e}{c}+\mathrm{H.c.})$. We move to an interaction picture with $\ke{\widetilde c}=\ke{c}$ and $\ke{\widetilde e}=e^{-i\omega_c t}\ke{e}$. We obtain
\begin{equation}
\widetilde H
=\hbar \Delta_c \kb{\widetilde e}{\widetilde e}
- \frac12 \left(\bm \mu_{ec} \cdot (\bm E_{c,0}+\bm E_{c,0}^*e^{2i\omega_c t})\kb{\widetilde e}{\widetilde c}+\mathrm{H.c.} \right)
\end{equation}
with the detuning $\Delta_c=\omega_{ce}-\omega_c$. Now, we neglect the rapidly rotating term $\bm E_{c,0}^*e^{2i\omega_c t}$. This is called rotating wave approximation. Obviously, the strength of the remaining atom-light coupling is described by the parameter $\Omega_c=\bm \mu_{ec} \cdot \bm E_{c,0}/\hbar$ which is called Rabi frequency. In the following we always work in this interaction picture but drop the tilde to simplify the notation. We obtain
\begin{equation}
H
=\hbar \Delta_c \kb{e}{e}
- \frac\hbar2 \left(\Omega_c \kb{e}{c}+\mathrm{H.c.} \right)
.
\end{equation}

\paragraph{Three-Level Atom.}

The probe light couples the atomic states $\ke{p}$ and $\ke{e}$ in Fig.\ \ref{fig-vSTIRAP}(a). A treatment similar to the control transition is applicable here. The corresponding detuning and Rabi frequency are $\Delta_p$ and $\Omega_p$. We use a matrix representation with respect to the basis $(\ke{p},\ke{c},\ke{e})$ and obtain \cite{fleischhauer:05}
\begin{equation}
H= -\frac\hbar2
\begin{pmatrix}
0 & 0& \Omega_p^* \\
0 & -2(\Delta_p-\Delta_c) & \Omega_c^* \\
\Omega_p & \Omega_c & -2\Delta_p \\
\end{pmatrix}
.
\end{equation}

\paragraph{Dark State.}

We assume that the two-photon Raman process connecting states $\ke{c}$ and $\ke{p}$ is resonant, i.e.\ $\Delta_p-\Delta_c=0$. As a result, the state
\begin{equation}
\label{dark-state}
\ke{D} \propto \Omega_c \ke{p} - \Omega_p \ke{c}
\end{equation}
is an eigenstate of $H$ with eigenvalue 0. It has no admixture of the excited state $\ke{e}$. So if spontaneous emission from the excited state is included in the model, an atom in this state will not show any spontaneous emission. This is why $\ke{D}$ is called a dark state. $\ke{D}$ has no admixture of $\ke{e}$ because the two amplitudes for exciting atoms to $\ke{e}$ from either $\ke{c}$ or $\ke{p}$ interfere completely destructively.

\paragraph{STIRAP.}

This dark state makes it possible to transfer all population adiabatically from one ground state to the other without causing spontaneous emission. To this end, start e.g.\ with all population initially in state $\ke{c}$. Turn on $\Omega_p$. Next, slowly increase $\Omega_c$ from zero to a nonzero value and (subsequently or simultaneously) slowly decrease $\Omega_p$ to zero. This will slowly rotate the dark state $\ke{D}$ from $\ke{c}$ to $\ke{p}$. If the rotation is slow enough, then -- by virtue of the adiabatic theorem of quantum mechanics -- all population will remain in the dark state $\ke{D}$ at all times, causing a passage of the population from $\ke{c}$ to $\ke{p}$. This is called stimulated Raman adiabatic passage (STIRAP). Finally, $\Omega_c$ can be turned off.

\paragraph{Vacuum STIRAP.}

The atom-cavity single-photon source that we discussed in Sec.\ \ref{sec-source} relies on STIRAP to suppress spontaneous emission. To explain this, we include the number of cavity photons in the notation, see Fig.\ \ref{fig-vSTIRAP}(b). The population is initially prepared in state $\ke{c,0}$. The control transition is $\lra{c,0}{e,0}$. The cavity field couples the transition $\lra{p,1}{e,0}$ with Rabi frequency $\Omega_p$.

It is customary to use $g=\Omega_p/2$ instead of $\Omega_p$. $g$ is obviously nonzero for the absorption process $\ke{p,1}\to\ke{e,0}$ because the atomic state $\ke{p}$ is exposed to the field of one photon. The fact that the Hamiltonian is Hermitian implies that the time-reversed process has the same Rabi frequency (assuming that $g$ is real). A rigorous treatment of the quantized light field confirms this result. Intuitively, it means that the vacuum field inside the cavity drives the emission $\ke{e,0}\to\ke{p,1}$. $g$ is sometime called single-photon Rabi frequency and sometimes vacuum Rabi frequency.

$g$ acts permanently, as long as the atom is inside the cavity. The single-photon generation process is the STIRAP sequence: Starting from zero, $\Omega_c$ is slowly increased to a value $\Omega_c\gg g$. This transfers almost all population into state $\ke{p,1}$ without spontaneous emission from state $\ke{e,0}$. The photon now present in the cavity mode will sooner or later be transmitted through the mirror, transferring the population into the final state $\ke{p,0}$. This is called vacuum STIRAP, because the emission into the cavity is driven by the vacuum field of the cavity.

If everything works adiabatically, the efficiency for emitting the photon by cavity transmission will reach 100\%. Among other things, this requires $g\gg \Gamma_e$, where $\Gamma_e$ is the spontaneous emission rate of state $\ke{e}$ into all directions other than the cavity mode. In present experiments the condition $g\gg \Gamma_e$ is not met well enough if one mirror is much more transparent than the other. This is the main reason why the efficiency in present experiments does not reach 100\%.

If the STIRAP process transfers the population to $\ke{p,1}$ much slower than a cavity decay time, then the population of state $\ke{p,1}$ will remain small at all times. Still, a large fraction of the population can be emitted by cavity decay. In this case, the subsystem $\ke{c,0},\ke{e,0},\ke{p,1}$ experiences loss from state $\ke{p,1}$ but the general adiabatic nature of the population transfer without population reaching state $\ke{e,0}$ remains intact. This is used for shaping the single-photon pulse.

\paragraph{Dark-State Polaritons in EIT and Slow Light.}

EIT and slow light are other effects that can be understood as adiabatic following in the dark state $\ke{D}$. Initially all atomic population is in state $\ke{p}$, corresponding to state $\ke{1}$ in Fig.\ \ref{fig-EIT-levels}. Control light is turned on. Subsequently, a pulse of probe light propagates through the atomic ensemble. The temporal variation of the envelope of the probe pulse is slow enough that the population adiabatically follows the dark state $\ke{D}$. When the probe pulse has left the medium, all atomic population is back in the initial state. At this time, the control light can be turned off. As the population is always in the dark state, no spontaneous emission occurs, yielding $\Im\chi=0$ at the two-photon resonance. Away from the two-photon resonance, however, no dark state exists, yielding $\Im\chi\neq0$, see Fig.\ \ref{fig-EIT-chi}(a).

As mentioned earlier, slow-light experiments must operate in the regime where the states $\ke{c}$ and $\ke{e}$ carry only a small fraction of the total atomic population at all times. The above discussion of STIRAP shows that the emergence of EIT at the two-photon resonance does not rely on this assumption. However, if the assumption is violated, no slow light will be obtained.

Generally, if light enters a medium, then the electric field will adopt a copropagating component of the induced dielectric polarization density $P=\epsilon_0 \chi E$. The corresponding quasi particle of the two co-moving components $P$ and $E$ in the medium is called polariton, rather than photon. In the context of EIT, this polariton follows a dark state and is therefore called dark-state polariton \cite{fleischhauer:00, fleischhauer:02}.

To understand slow and stopped light, we consider a probe pulse initially containing a single photon. With all atoms initially in atomic state $\ke{p}$, the dark-state polariton has the form
\begin{equation}
\label{dark-state-polariton}
\ke{D}\propto
\Omega_c \ke{1_\mathrm{photon}} \otimes \ke{\psi_\mathrm{initial}}
-\Omega_p \ke{0_\mathrm{photon}} \otimes \ke{\psi_\mathrm{magnon}}
\end{equation}
in analogy to Eq.\ \eqref{dark-state}. Here, $\ke{n_\mathrm{photon}}$ is a state with $n$ photons. $\ke{\psi_\mathrm{initial}}$ is the initial atomic state with all atoms in $\ke{p}$. $\ke{\psi_\mathrm{magnon}}$ is a single-magnon state in analogy to Eq.\ \eqref{magnon}. $\ke{D}$ becomes a pure photon in the limit $\Omega_c\gg\Omega_p$, whereas it becomes a pure magnon in the limit $\Omega_c\ll\Omega_p$.

The control light is on before the pulse enters the medium. Hence, initially $\Omega_c\neq0$ and $\Omega_p=0$. $\ke{D}$ is a pure photon. The pulse enters the medium and for small $\Omega_c$, it is adiabatically converted into an almost pure magnon. This suggests that the propagation speed of the polariton should be close to the propagation speed of the pure magnon. That speed is given by the differential photon recoil velocity, which is many orders of magnitude slower than the vacuum speed of light. This explains why the dark-state polariton is slow. It also explains why the group velocity is reduced when reducing $\Omega_c$. To create stopped light, one ramps $\Omega_c$ slowly all the way to zero with the probe pulse inside the ensemble and converts $\ke{D}$ into a pure magnon, which is almost at rest. At this point, one has completed a STIRAP process with full transfer with respect to the dark state polariton Eq.\ \eqref{dark-state-polariton} (but not with respect to the total atomic population). The readout process is the time-reversed version hereof.

\newpage


\end{document}